\documentclass{aa}
\usepackage{natbib}
\bibpunct{(}{)}{;}{a}{}{,}
\usepackage{graphicx}
\usepackage{txfonts}
\usepackage{hyperref}

\begin{document}

   \title{Determining the source and eruption dynamics of a stealth CME using NLFFF modelling and MHD simulations}

   \author{S.~L.~Yardley
          \inst{\ref{1},\ref{2}}
          \and
          P.~Pagano\inst{\ref{1},\ref{3},\ref{4}}
          \and
          D.~H.~Mackay\inst{\ref{1}}
          \and
          L.~A.~Upton\inst{\ref{5}}
          }

    \institute{School of Mathematics \& Statistics, University of St Andrews, North Haugh, St Andrews, Fife, KY16 9SS\email{sly3@st-andrews.ac.uk}\label{1} \and Mullard Space Science Laboratory, University College London, Holmbury St. Mary, RH5 6NT, UK\label{2}
    \and Dipartimento di Fisica \& Chimica, Università di Palermo, Piazza del Parlamento 1, I-90134 Palermo, Italy\label{3}
    \and INAF-Osservatorio Astronomico di Palermo, Piazza del Parlamento 1, I-90134 Palermo, Italy\label{4}
    \and Space Systems Research Corporation, 1940 Duke Street Suite 200, Alexandria, Virginia, 22314\label{5}
    }

   \date{Received ...}

 
  \abstract
   {Coronal mass ejections (CMEs) that exhibit weak or no eruption signatures in the low corona, known as stealth CMEs, are problematic as upon arrival at Earth they can lead to geomagnetic disturbances that were not predicted by space weather forecasters.}
   {We investigate the origin and eruption of a stealth event that occurred on 2015 January 3 that was responsible for a strong geomagnetic storm upon its arrival at Earth.}
   {To simulate the coronal magnetic field and plasma parameters of the eruption we use a coupled approach. This approach combines an evolutionary nonlinear force-free field model of the global corona with a MHD simulation.}
   {The combined simulation approach accurately reproduces the stealth event and suggests that sympathetic eruptions occur. In the combined simulation we found that three flux ropes form and then erupt. The first two flux ropes, which are connected to a large AR complex behind the east limb, erupt first producing two near-simultaneous CMEs. These CMEs are closely followed by a third, weaker flux rope eruption in the simulation that originated between the periphery of AR 12252 and the southern polar coronal hole. The third eruption coincides with a faint coronal dimming, which appears in the SDO/AIA 211~\AA\ observations, that is attributed as the source responsible for the stealth event and later the geomagnetic disturbance at 1 AU. The incorrect interpretation of the stealth event being linked to the occurrence of a single partial halo CME observed by LASCO/C2 is mainly due to the lack of STEREO observations being available at the time of the CMEs. The simulation also shows that the LASCO CME is not a single event but rather two near-simultaneous CMEs.}
   {These results show the significance of the coupled data-driven simulation approach in interpreting the eruption and that an operational L5 mission is crucial for space weather forecasting.}

               
    \titlerunning{Source and Eruption of a Stealth CME}

   \maketitle
%

\section{Introduction} \label{sec:intro}

Coronal mass ejections (CMEs) are large-scale eruptions of magnetised plasma that originate from the Sun and are considered to be the main drivers of space weather \citep{Gosling-1993, Hapgood-2011, Green-2015}. CMEs are usually associated with eruptive signatures observed low in the corona such as the rapid expansion of EUV loops, coronal dimmings, coronal waves, flares (including flare ribbons and arcades), and filament eruptions.

When these structures of magnetised plasma propagate outwards from the Sun they are detected in the coronagraph field-of-view (FOV) through the Thomson scattering of photons. CMEs observed in the plane-of-sky can exhibit a ``three-part'' structure that consists of a bright leading edge, a dark cavity, and a bright core of filament plasma \citep{Illing-1985}. Past observations have suggested that only 30~\% of CMEs possess this standard structure \citep{Webb-1987}. However, more recent studies that utilise multi-viewpoint observations taken by the STEREO spacecraft, have determined that at least 40~\% of CMEs have a three-part structure \citep{Vourlidas-2013, Vourlidas-2017}.

CMEs are the result of a storage-and-release process where magnetic stress and free magnetic energy are built up in the nonpotential coronal magnetic field. Despite decades of observations the exact physical processes involved in the occurrence of CMEs remain elusive. However, there are numerous models that exist to describe the formation and evolution of the pre-eruptive magnetic field configuration and the overlying coronal arcade. In these models, a flux rope, which comprises of twisted magnetic field, either exists prior to or forms during the eruption. If a flux rope exists prior to the eruption then a CME occurs due to a loss of equilibrium or an ideal instability such as the kink or torus instability \citep{Forbes-1991, Torok-2005, Kliem-2006, Kliem-2014}. Alternatively, the flux rope forms in-situ as a result of magnetic reconnection as described in the tether-cutting or breakout models \citep{Antiochos-1999, Moore-2001}. However, the pre-eruptive magnetic structure is most likely a hybrid configuration of sheared and twisted magnetic field where the nature of the configuration depends upon the stage of evolution of the pre-eruptive structure \citep[see the recent review by][]{Patsourakos-2020}. For in-depth reviews on the subject of CMEs, their formation and initiation, we refer the reader to \citet{Forbes-2000}, \citet{Forbes-2006}, \citet{Chen-2011}, \citet{Webb-2012}, \citet{Chen-2017}, \citet{Green-2018} and references therein.

When CMEs are Earth-directed they are the main source of intense geomagnetic storms. The interplanetary counterpart of CMEs (ICMEs), upon arrival in the near-Earth environment, can cause hazardous space weather effects, which severely impact our ground and space-based technological systems \citep{Schrijver-2015, Eastwood-2017}. A subset of ICMEs that have a magnetic flux rope configuration, known as magnetic clouds, can be particularly geoeffective \citep{Zhang-1988,Zhang-2004}. Magnetic clouds are geoffective as they can provide the strong, continuous southward-directed magnetic field component that is required for significant geomagnetic disturbances to occur \citep{Huttunen-2005, Zhang-2007, Gopalswamy-2008, Richardson-2013}. In-situ signatures of magnetic clouds include an enhanced magnetic field strength, a smooth, prolonged rotation of the magnetic field direction, and a reduction in proton temperature \citep{Burlaga-1981, Gonzalez-2011, Kilpua-2017}.

Stealth CMEs are solar eruptions that have no obvious visible signatures in the low corona \citep{Robbrecht-2009} however, are often associated with faint, slow CMEs observed in coronagraph data, and/or in-situ flux rope signatures. Stealth CMEs typically have velocities less than 300~km~s$^{-1}$, and generally originate from quiet Sun locations \citep{Ma-2010} or close to regions of open magnetic flux such as coronal holes \citep{DHuys-2014, Nitta-2017}. These events occur frequently for example, studies by \citet{Ma-2010} and \citet{Kilpua-2014} both examined the source regions of CMEs that were observed in coronagraph data during two different time periods during solar minimum in 2009. These studies found that 33~\% and 63~\% of their events were deemed to be stealth CMEs, respectively. The occurrence of stealth CMEs is therefore quite common however, their rate of occurrence is still an open topic of discussion.

Another key unanswered question is whether these `stealth' events are physically different from standard CMEs. A review of stealth CMEs by \citet{Howard-2013} suggests that they are part of a continuous spectrum of CMEs and their detection is limited by the capabilities of current instruments. In fact, most stealth CMEs can be associated with some form of low coronal signatures but additional processing techniques may be required to detect them \citep{Alzate-2017, Palmerio-2021}.

To help answer this question, we can use simulations to model the source region of the stealth CME. For example, \citet{Lynch-2016} performed a MHD simulation of the stealth CME reported by \citet{Robbrecht-2009} using the ARMS code \citep{DeVore-2008}. They were able to reproduce the observed white-light signatures of the CME, including the flux rope and three-part structure, and its propagation. Their results support the view that stealth CMEs are not fundamentally different from CMEs and form part of the lower range of the energy distribution of CMEs.

Despite stealth CMEs producing no obvious or very weak low coronal signatures and being faint, slow CMEs these events can still be geoeffective at Earth. \citet{Nitta-2017} investigated the origin of CMEs that showed no clear low coronal signatures, but were responsible for considerable disturbances at 1~AU during Solar Cycle 24. Out of 17 stealth events, six stealth CMEs produced moderate storms (minimum Disturbance storm time (Dst) $\leq$ $-$50~nT) and three produced strong storms (Dst $\leq$ $-$100~nT), highlighting the significance of these events for space weather forecasting. They found that coronal dimmings and post-eruption arcades are visible in the AIA observations only if long duration difference images of the source regions are constructed. The lack of the observational detection of stealth CMEs on the Sun is a huge problem for space weather forecasters as they cannot provide warnings of geoeffective events days in advance as required by many users.

In this paper, we investigate the stealth event of 2015 January 3 presented in the studies of \citet{Cid-2016} and \citet{Nitta-2017}. The stealth event led to an ICME that resulted in a strong geomagnetic storm upon its arrival at Earth. At the time of the stealth event there were several possible sources present on the Sun that could be responsible. Therefore, to analyse the source and eruption of the stealth event, we couple the global evolutionary nonlinear force-free field (NLFFF) model of \citet{Mackay-2006a}, which can simulate the solar corona over periods of months to years, to the global MHD simulations of \citet{Pagano-2018}.

The paper is organised as follows. Section~\ref{sec:obs} describes the remote and in-situ observations of the occurrence of the stealth event and its possible CME/ICME counterparts. Section~\ref{NLFFFEvolutionaryModel} describes the NLFFF evolutionary model used to simulate the coronal magnetic field evolution and its results. Section~\ref{coupleMHD} explains the coupling of the NLFFF evolutionary model with the MHD simulation and shows the evolution of multiple eruptions in the MHD simulation. By using the MHD simulation the origin of the stealth event is determined. Finally, Sect.~\ref{sec:sum} gives the conclusions and a discussion.


\section{Observations} \label{sec:obs}

We will now briefly describe the observations acquired from the multiple space-based instruments that we use in this study. The in-situ data are recorded by several spacecraft positioned in the near-Earth environment at the Lagrange 1 (L1) point. The magnetic field data are taken by the Magnetometer instrument (MAG) on board the {\it Advanced Composition Explorer} (ACE; \citealt{Smith-1998}) while the plasma parameters and Dst index are provided by OMNI \citep{King-2005}. The time cadence of the data ranges from 16~s to 1~hr.

The occurrence of CMEs are monitored using the coronagraph observations taken by the Large Angle and Spectrometric Coronograph (LASCO; \citealt{Brueckner-1995}) on board the {\it Solar and Heliospheric Observatory} (SOHO; \citealt{Domingo-1995}). Unfortunately, there are no STEREO observations available during the time period of our study. The photospheric magnetic field evolution is analysed using full disk line-of-sight (LoS) magnetogram data taken from the Helioseismic Magnetic Imager (HMI; \citealt{Schou-2012}) on board the {\it Solar Dynamics Observatory} (SDO; \citealt{Pesnell-2012}). The coronal evolution is studied using the Atmospheric Imaging Assembly (AIA; \citealt{Lemen-2012}) also on board SDO. The AIA instrument provides observations of multi-thermal plasma, which allows us to analyse the coronal response to changes in the photospheric magnetic field. In particular, we have analysed the 171, 193 and 211\,\AA\ wavebands.

\subsection{In-situ observations}

\begin{figure*}[h]
\centering
\includegraphics[width=1\textwidth]{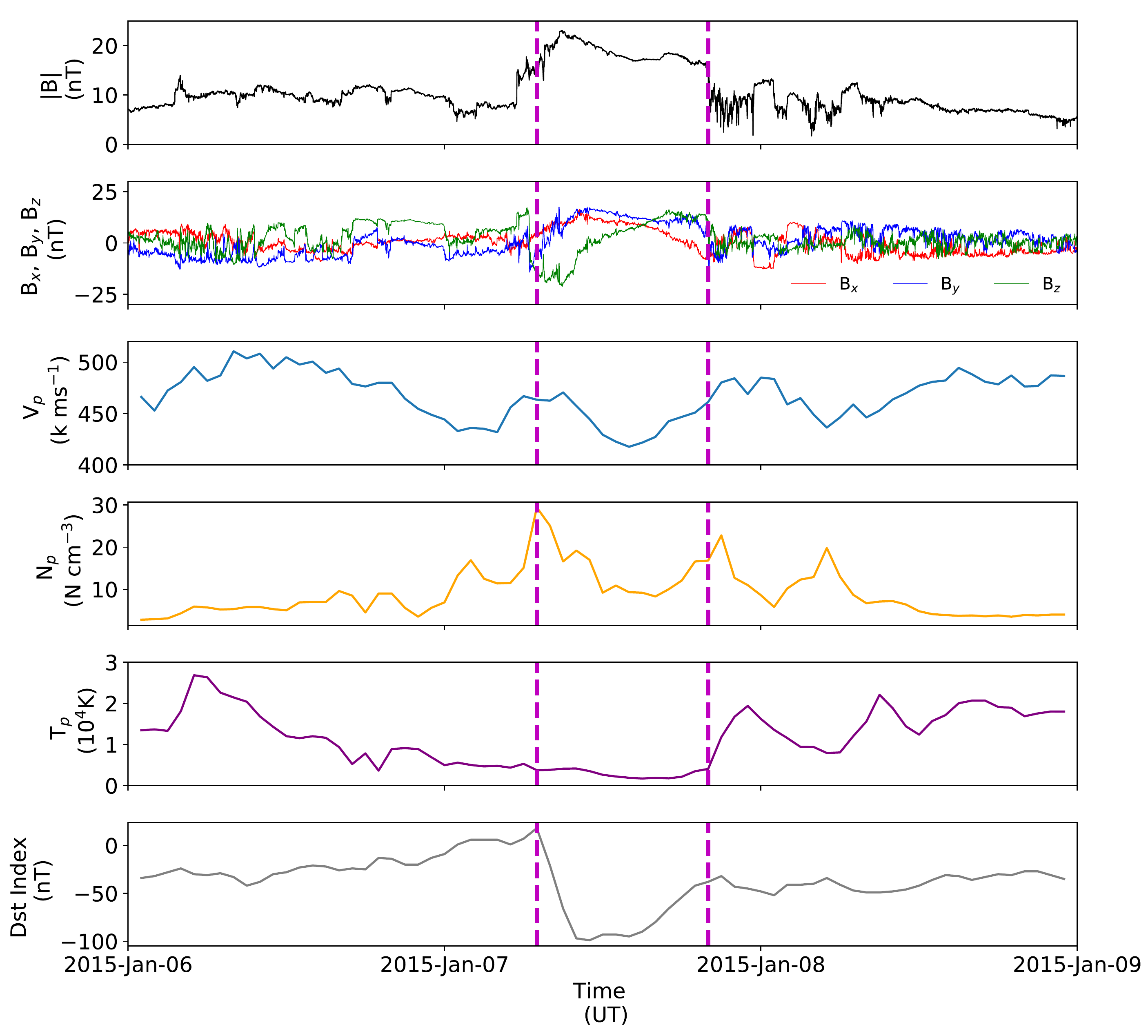}
\caption{In-situ data taken between 2015 January 6--9, which shows the presence of an ICME on 2015 January 7. The panels from top to bottom show the magnitude of the magnetic field (ACE 16~s data), the magnetic field components (B$_{x}$, B$_{y}$, B$_{z}$) in Geocentric Solar Ecliptic (GSE) coordinates (ACE 16~s data), the solar wind speed (V$_{p}$; OMNI 1~hr data), the solar wind density (N$_{p}$; OMNI 1~hr data), the proton temperature (T$_{p}$; OMNI 1~hr data), and the Dst index (OMNI 1~hr data). The purple dashed lines indicate the ICME interval from to 07:00~UT to 20:00~UT taken from the Richardson and Cane ICME catalog (\url{http://www.srl.caltech.edu/ACE/ASC/DATA/level3/icmetable2.htm\#(e)~}). \label{fig:fig1}}
\end{figure*}

On 2015 January 7, between the hours of 06-12 UT, the NOAA Space Weather Prediction Center recorded a maximum planetary K-index (Kp) 7 and hence issued an alert for a severe (G3-strong) geomagnetic storm. Figure~\ref{fig:fig1} shows the disturbance detected in the magnetic field data, plasma data and Dst Index due to the arrival of an ICME at 1~AU. The disturbance detected is due to an ICME that arrives at L1 on 2015 January 7 with a duration of $\sim$13~hrs. The start and end times of the ICME are given by the purple dashed lines in Fig.~\ref{fig:fig1} and are taken from the Richardson and Cane online ICME catalogue. The ICME is classified as a magnetic cloud due to the enhancement in the magnetic field strength, the smooth rotation of the $B_{z}$ component of the magnetic field, and a decrease in the proton temperature. The rotation of the $B_{z}$ component from south to north and positive $B_{x}$, $B_{y}$ suggests that the magnetic field is left-handed or has negative helicity, as stated in \citet{Cid-2016}. The minimum Dst index recorded during this interval was $-$99~nT corresponding to a moderate geomagnetic storm.
Prior to the disturbance, the proton density is depleted and the proton temperature is high suggesting the arrival of solar wind from the southern polar coronal hole. However, the moderate velocity of the wind (500~km~s$^{-1}$) suggests that the wind originates close to the coronal hole boundary.
The geomagnetic disturbance was unexpected as no Earth-directed CMEs were observed and solar activity was at a low level during this time period. At the time of the storm the only identifiable source of the disturbance was the CME observed in LASCO/C2 on January 3.
This CME was initially believed to originate from the farside of the Sun as no solar eruptions were observed on disk. In the present paper we will show that this CME was not responsible for the geomagnetic storm.

\subsection{Photospheric and coronal evolution}

\begin{figure}
\resizebox{\hsize}{!}{\includegraphics{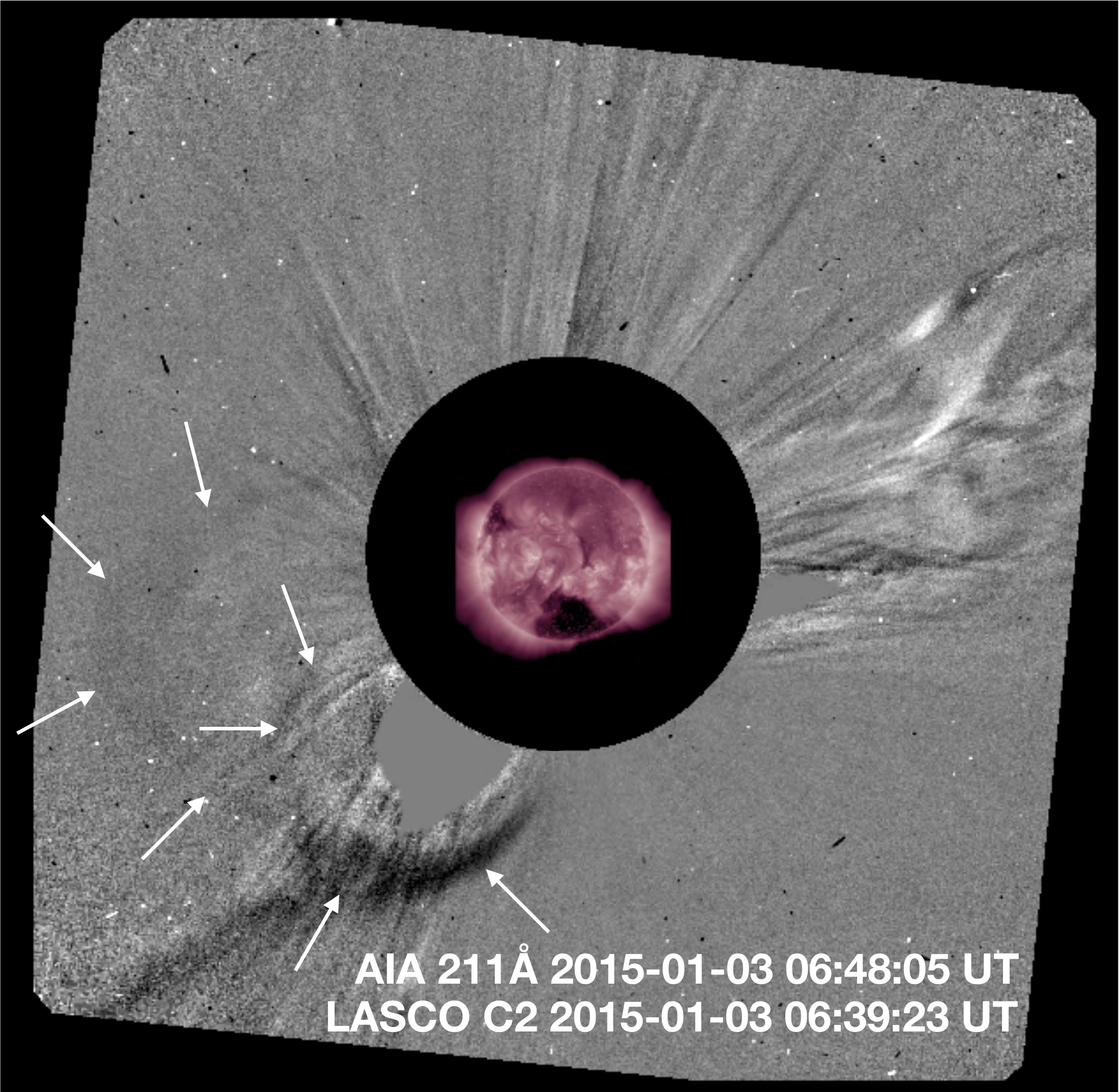}}
\caption{Running difference image taken by LASCO/C2, where the leading edge and the double-lobed structure of the CME is highlighted by white arrows. The corresponding SDO/AIA 211\,\AA\ image is shown at the centre. \label{fig:fig2} }
\end{figure}

The only easily identifiable CME that could be responsible for the disturbance at 1 AU on 2015 January 7 is a partial halo CME that is detected by LASCO/C2 on 2015 January 3 at 03:12~UT. This CME is observed in LASCO/C2 at a position angle of 118$^{\circ}$ with an angular with of 153$^{\circ}$. It is a slow CME with a diffuse front (white arrows in Fig.~\ref{fig:fig2}) travelling at 163~km~s$^{-1}$. The CME is difficult to track in the observations due to its faint emission and also the presence of a helmet streamer, which appears saturated in the difference images. The CME is recorded as a single partial halo CME however, on close inspection of the LASCO observations (Fig.~\ref{fig:fig2}), the CME has a double-lobed structure. This raises the question as to whether a single partial halo CME or two near-simultaneous CMEs occurred. To determine the source region of the partial halo CME observed by LASCO we analyse the photospheric and coronal observations taken by SDO/AIA and HMI.

\begin{figure*}[h]
\centering
\includegraphics[width=1.0\textwidth]{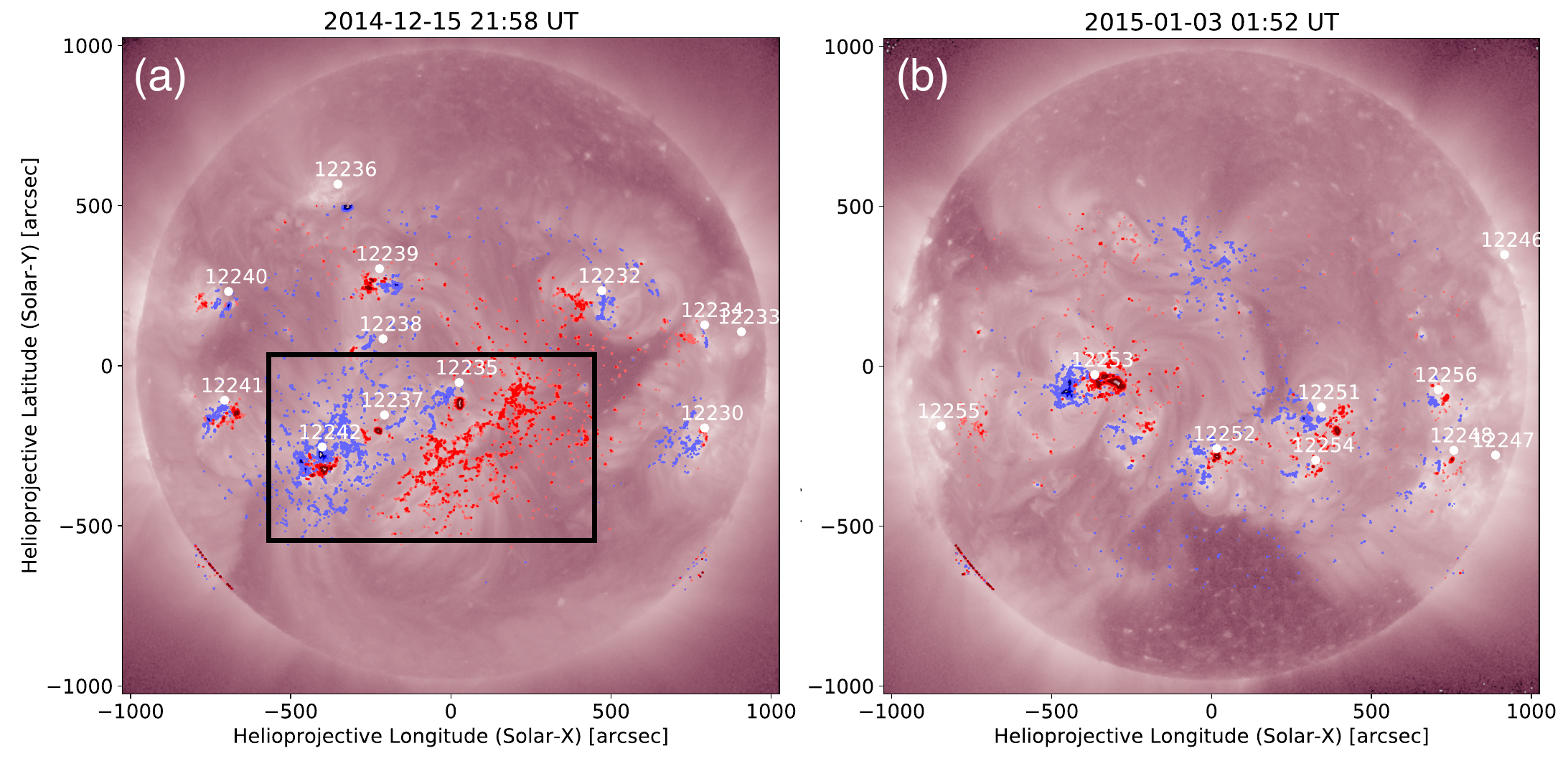}
\caption{Full disk images of the Sun in the 211\,\AA\ waveband taken by SDO/AIA with the corresponding photospheric magnetic field overlaid from SDO/HMI. Panel (a) shows the coronal emission and surface magnetic field near the beginning of CR 2158, where the black box highlights a large AR complex containing ARs 12235, 12237 and 12242. The image in panel (b) is taken at the end of CR 2158 after the large AR complex has rotated behind the limb and an eruption occurs close to the southern polar coronal hole. Red (blue) contours represent positive (negative) magnetic field values saturated at $\pm$500~G. \label{fig:fig3} }
\end{figure*}

The CME occurs during solar maximum in Carrington Rotation (CR) 2158 when the structure of the Sun's magnetic field is very complex. CR 2158 begins on 2014 December 8 and ends on 2015 January 4. Near the beginning of this rotation on 2014 December 15, 19 days before the CME, there are 12 active regions (ARs) observed on the front side of the solar disk (panel (a) of Fig.~\ref{fig:fig3}). These ARs are evenly distributed across the northern and southern hemispheres. The ARs that are located in the southern hemisphere exhibit significant flaring activity and multiple eruptions during their disk passage. In particular, there is a large AR complex (black box in panel (a) of Fig.~\ref{fig:fig3}) that is the main source of activity during this period. The AR complex, which spans almost half of the southern hemisphere, is composed of ARs 12235, 12237, and 12242 that are embedded in decayed quiet Sun magnetic field. Due to the size of this AR complex, it most likely survives as it rotates around to the farside of the solar disk, where it is located behind the east limb of the Sun at the end of CR 2158. In fact, during the next rotation (CR 2159), the large AR complex is visible on the disk and is composed of AR 12259, 12261 and a large area of decayed positive magnetic field.

\begin{figure*}[h]
\centering
\includegraphics[width=1.0\textwidth]{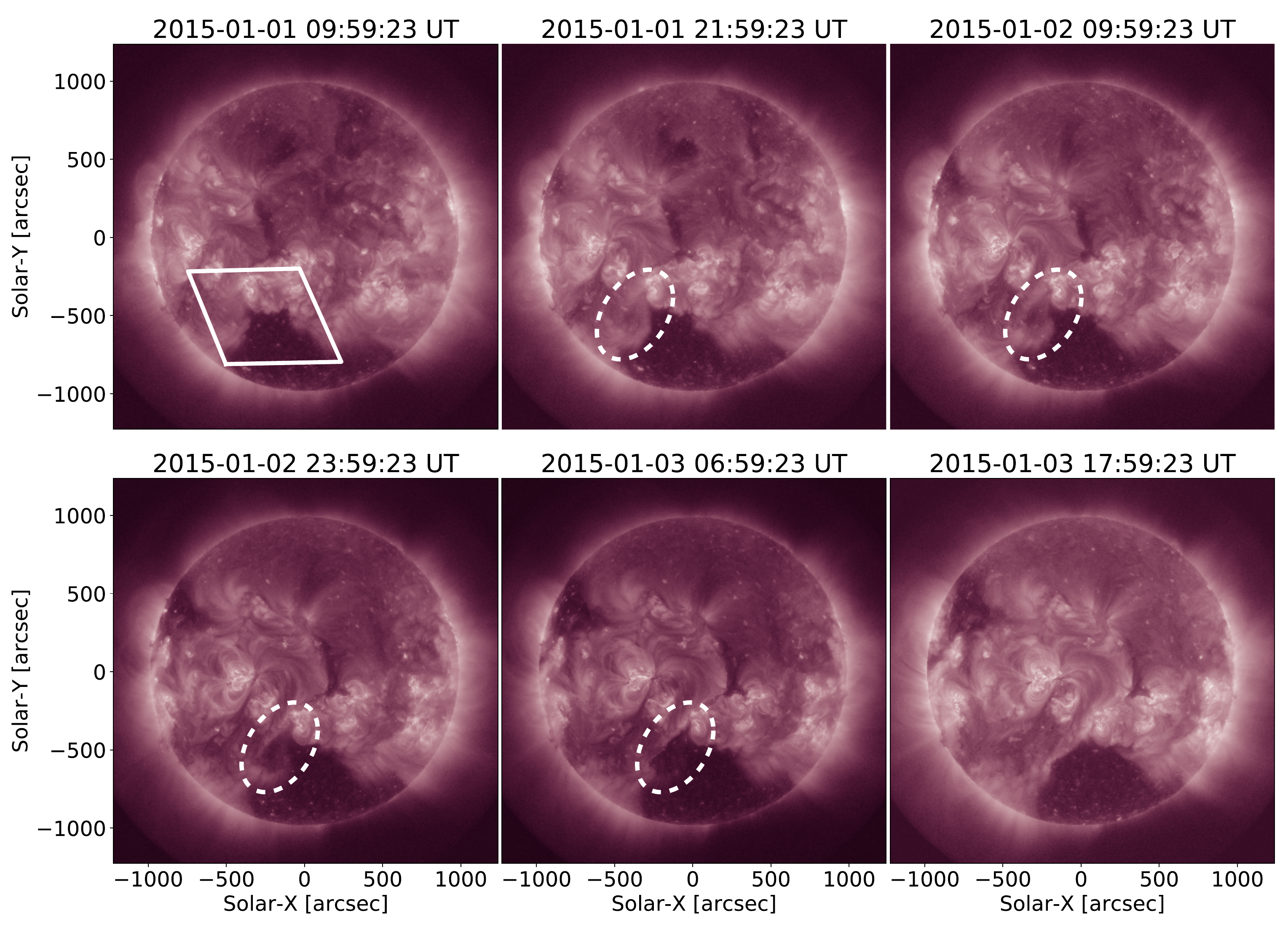}
\caption{Sequence of 211\,\AA\ full disk images taken by SDO/AIA, which shows a weak coronal dimming in the southern hemisphere. This low coronal eruption signature is the only observed feature associated with the stealth event. It is indicated by the white dashed ellipse in the central four panels. The parallelogram in panel (a) indicates the FOV of the online movie that shows the stealth event. \label{fig:fig4} }
\end{figure*}

On 2015 January 3, at the end of CR 2158, a CME occurs, which is assumed to be responsible for the geomagnetic disturbance on January 7. At the time of the CME there are several possible source regions visible on the solar disk in the southern hemisphere (see panel (b) of Fig.~\ref{fig:fig3}), including 7 ARs and a large, extended polar coronal hole. \citet{Cid-2016} discusses the several possible sources that could be responsible for the ICME and the associated CME. The possible sources include multiple AR filaments located in ARs 12251, 12253 and 12254 and a filamentary structure located to the east of the southern polar coronal hole \citep[see Fig. 3 of][]{Cid-2016}. However, \citet{Cid-2016} comes to no definite conclusions as to which of these sources is responsible. \citet{Nitta-2017} suggest that a faint coronal dimming that is visible in the SDO/AIA 211\,\AA\ observations (dashed ellipse in Fig.~\ref{fig:fig4}) is the source region of the CME. The dimming is most evident in difference images that have a long temporal separation \citep[see Fig. 13 of][]{Nitta-2017}. The coronal dimming is first visible on 2015 January 1 around $\sim$22:00~UT and starts to develop during January 2. The dimming appears to be connected to the eastern periphery of AR 12252, a relatively simple AR. The structure evolves and the dimming intensifies at the beginning of January 3. During this time, the dimming structure rapidly expands and merges into the extended southern polar coronal hole.

The observations from LASCO/C2 and AIA suggest that a stealth event may have occurred. The stealth event seems to originate from the region located between the east periphery of AR 12252 and the southern polar coronal hole. The event is stealthy as only weak eruption signatures, in the form of a faint dimming, are observed in the low corona. The relationship between this stealth event and the LASCO/C2 CME is however unclear, even though they occur at approximately the same time. Also, it is unclear how the CME relates to the ICME. From the interpretation of the observations there were no strong signatures of a solar eruption on the disk. Due to this, it was initially determined that the CME originated from the farside of the Sun and an alert was not issued by space weather forecasters.

There are several questions, regarding the stealth event, that remain unanswered and cannot be resolved from the analysis of the observations alone. Firstly, it is unclear whether the single partial halo observed by LASCO is responsible for the ICME that caused the geomagnetic disturbance. Secondly, we can not determine whether the CME originates  from the source region located close to  the coronal hole or from a region on the farside of the Sun. It is also impossible to distinguish whether the CME is a single partial halo CME or two CMEs that erupt almost simultaneously. To answer these questions and to determine the source and dynamics of the observed eruptions we apply a combined simulation approach by coupling a global NLFFF model to a MHD simulation. We describe the NLFFF model and the MHD simulation in the Sections below.

\section{NLFFF evolutionary model}
\label{NLFFFEvolutionaryModel}

To investigate the source region and occurrence of the stealth event on 2015 January 3 we use a global evolutionary magnetofrictional model. This model allows us to study the build-up of stress and energy in the magnetic field during the weeks and months prior to the eruption of the stealth event. The global evolutionary model has been originally developed by \citet{vB-2000} and \citet{Mackay-2006a, Mackay-2006b} and applied successfully on numerous occasions to simulate the evolution of the coronal magnetic field \citep{Yeates-2008a, Yeates-2008b, Yeates-2009a, Yeates-2009b, Yeates-2012, Mackay-2014, Mackay-2018, Yeates-2018}.

\subsection{Initial condition \& method}

To simulate the occurrence of the stealth event an initial field extrapolation was constructed from a synoptic magnetogram taken on 2014 September 1. Once corrected for differential rotation the initial condition was evolved forward in time to simulate the photospheric and coronal magnetic field at future times. One important aspect of the photospheric simulation is that to maintain their accuracy new magnetic flux must be inserted to represent the emergence of new bipoles. The emergence of new magnetic bipoles is derived from the Advective Flux Transport (AFT) model \citep{Upton-2014a,Upton-2014b, Ugarte-Urra-2015}.
In total, 197 magnetic bipoles were inserted into the simulation, which runs for 200 days from 2014 September 1 until 2015 March 20, the date of a total solar eclipse. It takes the simulation around 6-8 weeks (ramp-up phase) to lose the memory of the initial potential field configuration and for non-potentiality to be built up in the coronal magnetic field. 

The AFT model uses the line-of-sight magnetic field from HMI magnetograms to simulate the magnetic field evolution as a result of differential rotation, meridional circulation and small-scale convective flows. The location and properties of the new bipoles are then determined from the AFT simulations through the application of a three-stage semi-automated procedure (see Appendix B of \citealt{Yeates-2018}). First, successive magnetograms are compared every 24 hours to identify new magnetic bipoles. Second, the properties such as the longitude, latitude, separation of the magnetic polarities, magnetic flux, and tilt angle of the identified bipoles are computed using 1~hr data from the AFT model over a time period of a few days. This includes the determination of the maximum magnetic flux of the bipole. The bipoles are then introduced into the NLFFF simulations mathematically as idealised magnetic bipoles at the time the bipole reaches its total maximum flux.
These bipoles are inserted into the pre-existing field as isolated flux and can contain either zero, negative or positive self-helicity. Once inserted, the coronal field of the bipole relaxes towards a force-free state due to magnetofriction. In addition, the inclusion of $\eta {\mathbf j}$
diffusion in the corona allows the bipole to reconnect with the overlying 
coronal field of the surrounding region.

In the NLFFF model, the driving of the photospheric boundary due to flux transport processes causes the coronal field to diverge from a force-free equilibrium. To return the coronal field to a force-free state we relax the field using the magnetofrictional relaxation technique of \citet{Yang-1986}. The relaxation is achieved by using an artificial friction term in the induction equation that sets the plasma velocity to be proportional to the Lorentz force. In the simulation the photospheric driving and coronal relaxation are performed simultaneously. Over the entire 6 month period an outflow velocity term is also prescribed, which ensures that the coronal field is radial at the location of the source surface (r = 2.5 R$_{\odot}$). Two reasons for including this outflow are to allow erupting flux ropes to be removed from the coronal volume and to simulate the effect of the solar wind along open field lines. The result of the simulation is the generation of a continuous time sequence of NLFFFs over long time scales. This particular method allows for the retention of memory of the previous magnetic field configuration as well as the build-up and transport of free magnetic energy and helicity in the simulation. A full description of the computational grid and boundary conditions used in the simulation can be found in \citet{Mackay-2014}.

\subsection{NLFFF magnetic field evolution}
We now present the results of the NLFFF evolutionary model that will be used to produce a nonpotential initial condition for the MHD simulations in the next Section. The NLFFF simulation can be run with numerous additional parameters to find a best-fit model that can reproduce the observations most accurately. In particular to constrain the model parameters, the locations of magnetic flux ropes in the NLFFF model can be compared to H$\alpha$ observations of filament channels taken on the day of the eclipse (2015 March 20) to determine the goodness-of-fit. The NLFFF simulations used in this paper to produce the initial condition in the MHD simulations are nearly identical to the `mf - evolving magnetofrictional'
simulation found in the review paper of \cite{Yeates-2018}. The simulation gave the best-fit to the H$\alpha$ observations when the magnetic bipoles were inserted with a twist parameter of $\pm$0.4 with positive (negative) twist assigned to bipoles that emerged in the southern (northern) hemisphere. In addition, Ohmic diffusion was included in the induction equation with a resistive coefficient of 60~km$^{2}$~s$^{-1}$. This NLFFF simulation was previously used in \cite{Meyer-2020} to compare the large-scale structure of the magnetic field from the simulations with off-limb SWAP images. While nearly identical, the simulation presented in the present paper is slightly different from those previously published as in the few days prior to the eruption we deviated from a default value of $\pm$0.4 for the newly emerging bipoles and assigned a zero twist for ARs 12252 and 12253 to be more consistent with the untwisted nature of the bipoles in the observations. 

\begin{figure*}[h]
\centering
\includegraphics[width=1.0\textwidth]{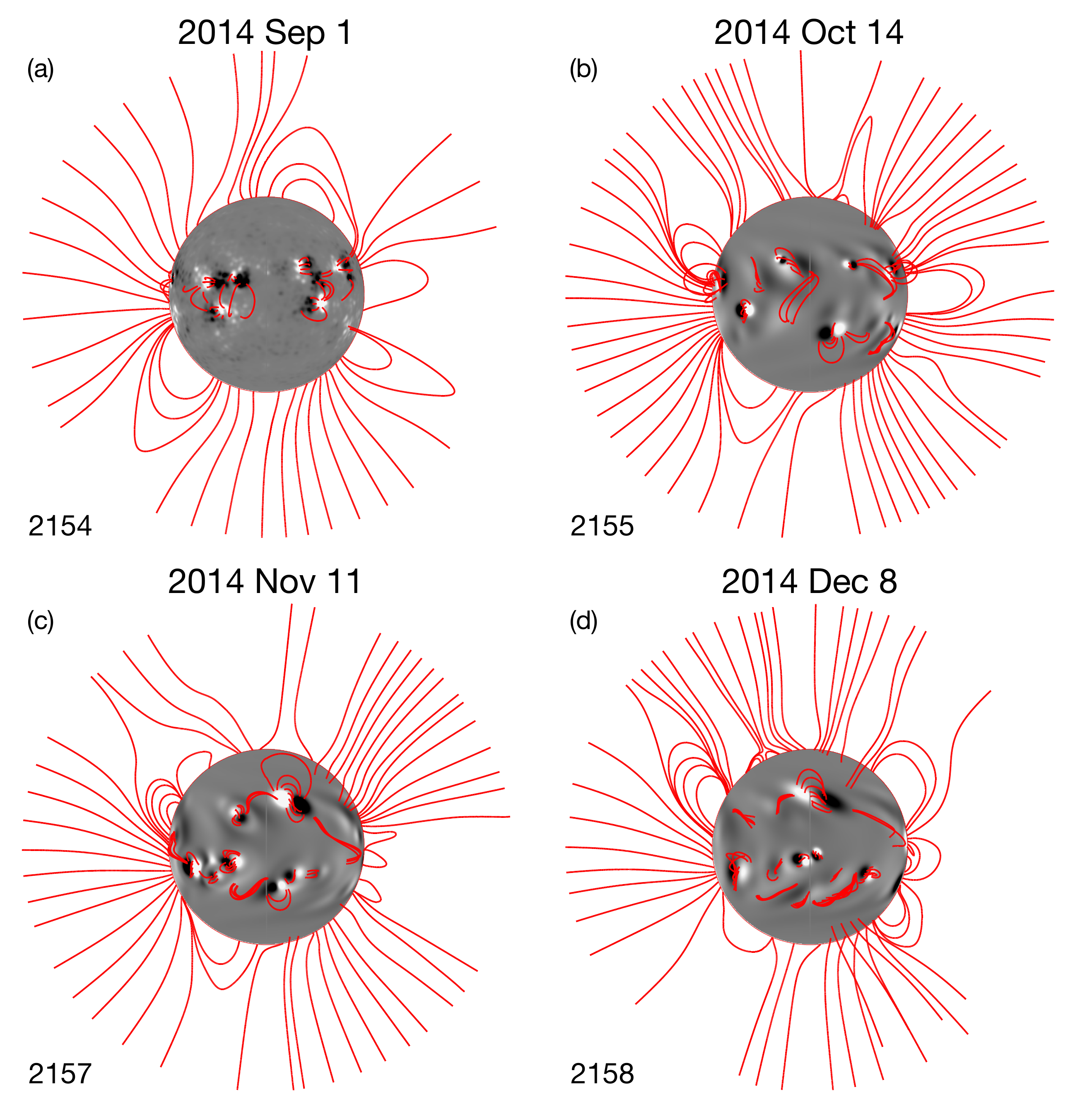}
\caption{Magnetic field structure taken from the global NLFFF evolutionary model in the time period prior to the eruption of the stealth event that occurred on 2015 January 3. The panels show the radial magnetic field of the photosphere (grey) where white (black) regions correspond to positive (negative) magnetic field. The red lines are representative magnetic field lines from the model to show the configuration of the coronal magnetic field. The panels are plotted in a co-rotating frame at the Carrington rate where the corresponding Carrington Rotation is given in the bottom left. In panel (a), the initial potential condition (day 1) of the model is shown where central meridian is at 215$^{\circ}$ longitude. For panels (b)--(d) central meridian is at 0$^{\circ}$ longitude. \label{fig:fig5} }
\end{figure*}

Figure~\ref{fig:fig5} shows the photospheric and coronal magnetic field evolution taken from the global NLFFF evolutionary model during the time leading up to the stealth event. The initial condition of the simulation was a potential field extrapolation constructed on 2014 September 1, which is shown in panel (a) of Fig.~\ref{fig:fig5}. The stealth event is observed to occur on 2015 January 3, which corresponds to day 123 of the simulation. As the ramp-up phase of the simulation is approximately 6-8 weeks the magnetic field is fully nonpotential before the stealth eruption occurs. This is evident by the build-up of sheared and twisted coronal magnetic field in the computational domain as displayed by Fig.~\ref{fig:fig5}.

\begin{figure*}[h]
\centering
\includegraphics[width=0.7\textwidth]{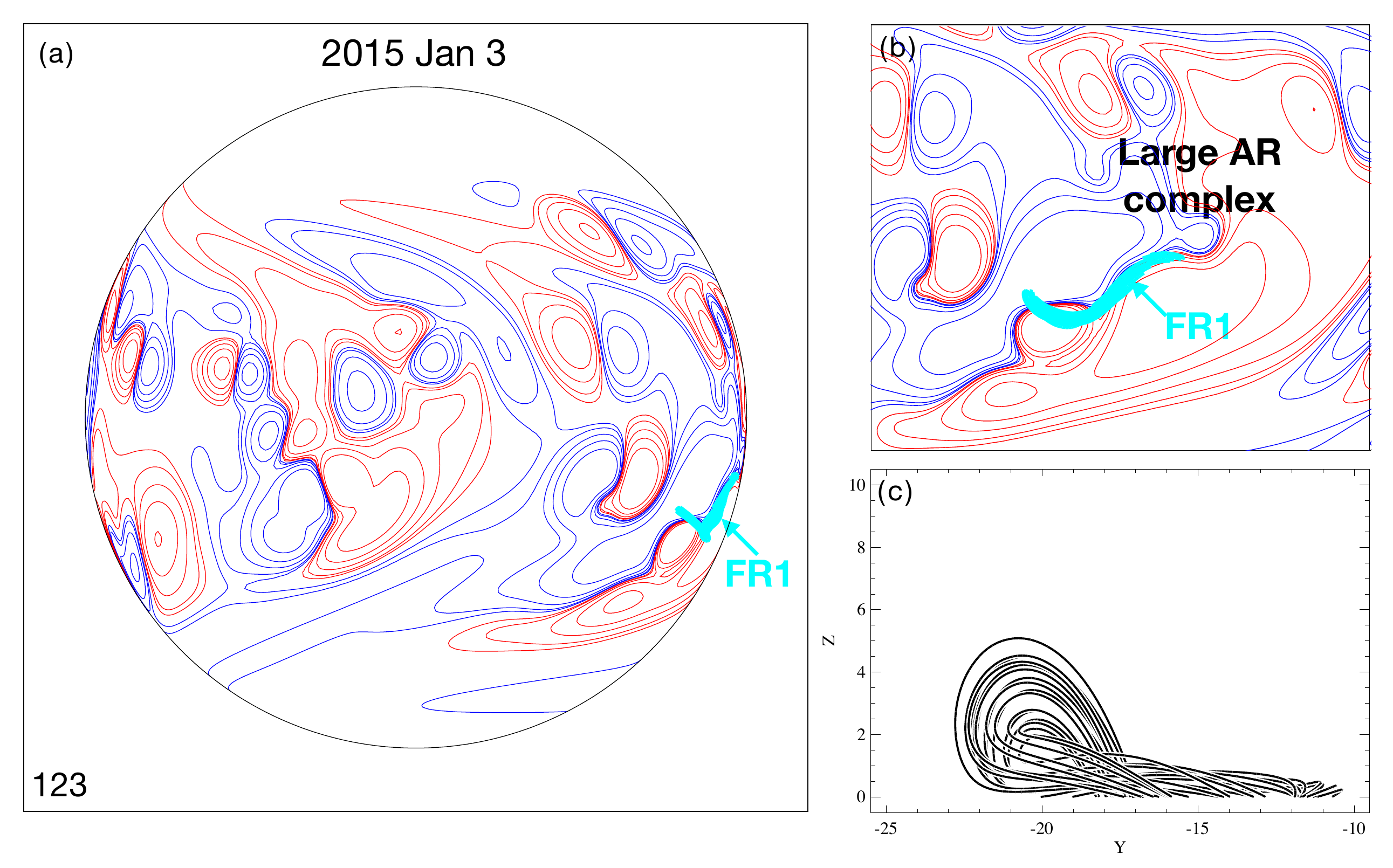}
\caption{Magnetic field lines taken from the global NLFFF evolutionary model that show the presence of a flux rope (FR1) along the PIL of the large AR complex that is located behind the east solar limb. The panels are plotted using the same coordinate frame as Fig.~\ref{fig:fig5} and the day of the simulation is given in the bottom left of panel (a). The blue (red) contours represent negative (positive) photospheric magnetic field. Panel (a) shows the entire solar disk as it would be viewed from the farside of the Sun from the L3 perspective with a central meridian longitude of 180$^{\circ}$ and representative field lines of FR1 (light blue). Panels (b) and (c) show a close up of the flux rope and the axis of the flux rope in the y-z plane, respectively. \label{fig:fig6} }
\end{figure*}

\begin{figure*}[h]
\centering
\includegraphics[width=0.7\textwidth]{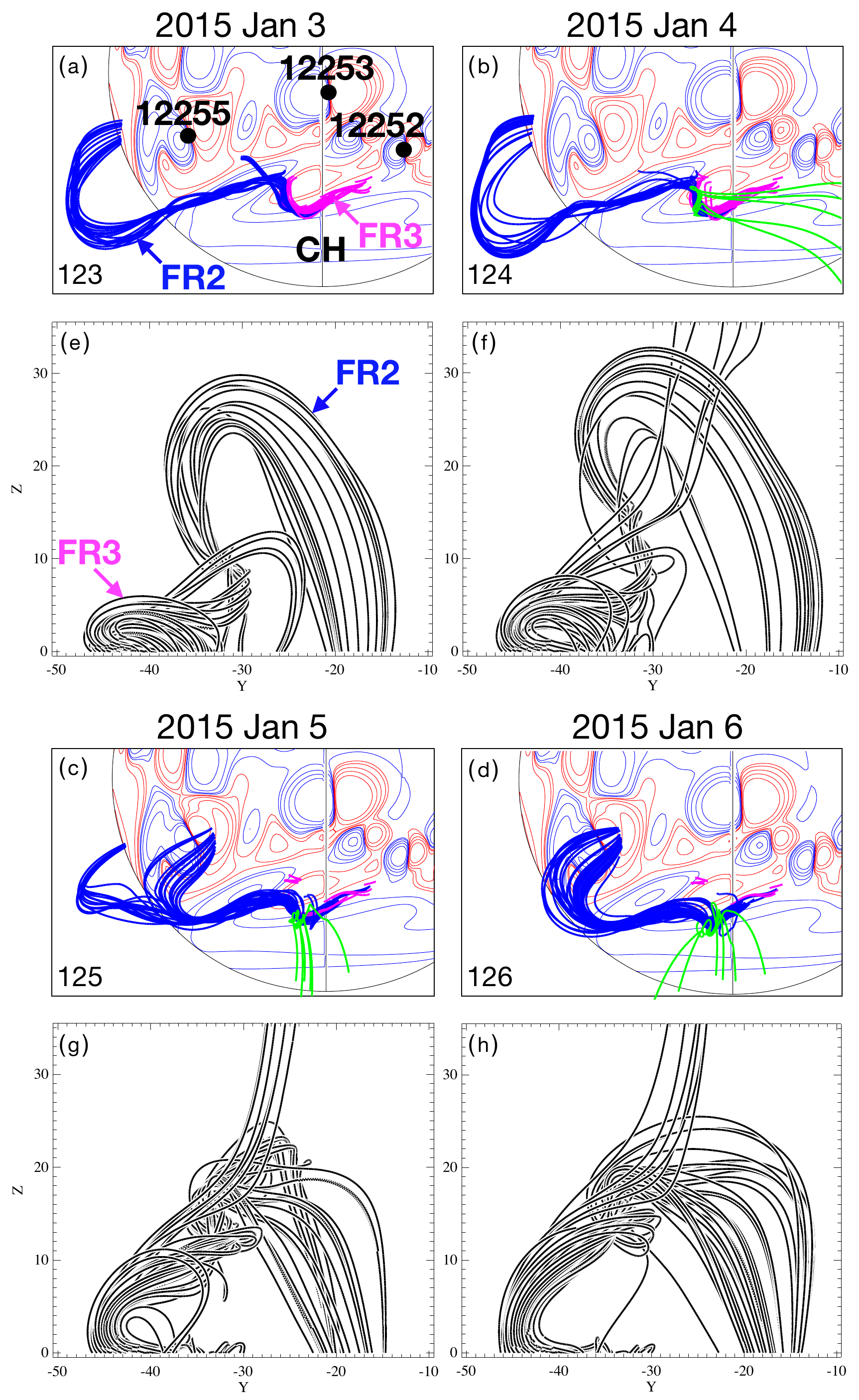}
\caption{Magnetic field lines taken from the global NLFFF evolutionary model showing the configuration of the magnetic field in the time period surrounding the eruption of the stealth event. Panels (a)--(d) show representative magnetic field lines of two flux ropes denoted by FR2 (blue lines) and FR3 (magenta lines), and also open field (green lines). These panels are plotted in the same co-rotating frame as Fig.~\ref{fig:fig5} and the day of the simulation is given in the bottom left. For panels (a)--(d) the central meridian longitude is 13$^{\circ}$, 0$^{\circ}$, 347$^{\circ}$ and 334$^{\circ}$, respectively. The blue (red) contours represent negative (positive) photospheric magnetic flux and the locations of ARs 12252, 12253, 12255, and the extended south pole coronal hole (CH) are given in the first panel. The same magnetic field lines are shown in the y-z plane in panels (e)--(h). \label{fig:fig7} }
\end{figure*}

Figure~\ref{fig:fig6} shows the surface magnetic flux and magnetic field lines taken from the NLFFF evolutionary model on the day of the stealth event as seen from the L3 perspective i.e. the farside of the Sun. In contrast, Figure~\ref{fig:fig7} shows the surface magnetic flux and magnetic field lines taken from the NLFFF evolutionary model on the days surrounding the time of the stealth event (2015 Jan 3--6) from the L1 perspective. Prior to the stealth event, three flux ropes have formed in the NLFFF model that we will now refer to as FR1, FR2, and FR3. FR1, FR2, and FR3 correspond to the light blue, dark blue, and magenta flux ropes shown in Figures~\ref{fig:fig6} and \ref{fig:fig7}.
The first flux rope (FR1 in Fig.~\ref{fig:fig6}) has formed along the PIL of the large AR complex that is located behind the east limb. The second flux rope (FR2 in Fig.~\ref{fig:fig7}) has footpoints rooted in the positive photospheric magnetic field of the large AR complex located on the east limb and negative polarity quiet Sun magnetic field that is close to the extended polar coronal hole in the southern hemisphere. The third flux rope (FR3 in Fig.~\ref{fig:fig7}) is located between the eastern periphery of AR 12252, the polar coronal hole, and a region of positive quiet Sun magnetic flux, which is adjacent to the western footpoints of FR2. The location of FR3 matches the location of the faint coronal dimming that is observed to occur in the observations (see Sect.~\ref{sec:obs}). All three flux ropes have formed mainly as a result of differential rotation and surface diffusion that leads to convergence and cancellation, followed by magnetic reconnection.

In the NLFFF evolutionary model, in the days leading up to the stealth event and CME on 2015 January 3, reconnection occurs between the footpoints of the two flux ropes FR2 and FR3. The reconnection continues in the days subsequent to the stealth event and eventually a new flux rope is formed, which is a combination of FR2 and FR3. At the same time, the overlying magnetic field of the two flux ropes opens up (green magnetic field lines in panel (b) of Fig.~\ref{fig:fig7}), which allows the newly formed flux rope to rise in the domain (see panels (g) and (h) of Fig.~\ref{fig:fig7}). The occurrence of external reconnection above the magnetic flux rope and the resulting rising motion of the flux rope suggests that a loss of equilibrium has occurred and the simulation has successfully captured the pre-eruptive magnetic configuration of the structure that results in the stealth event. It also suggests that, due to the interaction between the two magnetic structures, there is a connection between the large AR complex and the polar coronal hole. 

The loss of equilibrium of the flux rope (formed from FR2 and FR3) in the NLFFF evolutionary model occurs after the time of the stealth event in the observations and extends over a longer time period. This can occur due to a combination of reasons. Firstly, the build-up of energy in the nonpotential coronal field in the model occurs over a long timescale of months therefore, we do not expect the eruption to occur in the NLFFF model at the exact time as in the observations. Although, we would expect the eruption to occur within a few days. The fact that the eruption is reproduced within a few days of it occurring in the observations is a positive result. Secondly, we injected all of the bipoles with a twist parameter of magnitude 0.4 until a few days prior to the stealth event. On the Sun, each bipole will have a slightly different twist, which could alter the time of the eruption. Finally, the longer duration of the processes in the NLFFF simulations is due to the relaxation approach that dampens the dynamics.

While the magnetofrictional approach is computationally efficient and provides the pre-eruptive magnetic field configuration in the build-up to eruption over long time scales, to simulate the complete dynamics of eruptions a full MHD simulation is required. We therefore take the global magnetic field from the NLFFF evolutionary model, on the day of the stealth event (day 123, which corresponds to 2015 January 3), and use it as the initial condition of a MHD simulation. This will allow us to consider the true eruption dynamics of the flux ropes. This is described in the next Section, where we also consider the eruption of FR1, FR2, and FR3 in the MHD simulation.

\section{Global MHD simulation}
\label{coupleMHD}

The NLFFF evolutionary model described in Section~\ref{NLFFFEvolutionaryModel} accurately reproduces the build-up of magnetic stress and energy in the corona over long timescales such as days or months. However, in order to model the dynamics of solar eruptions on shorter timescales such as minutes or hours we require full MHD. The two approaches can be coupled by using the magnetic field configuration of the global corona from the NLFFF evolutionary model as the initial condition in a MHD simulation. Thus providing a realistic data-driven initial condition to model the eruption. This combined approach of coupling the two methods has been successfully adopted in a series of studies focused on advancing the progress of the data-driven modelling on erupting time scales \citep{Pagano-2013a,Rodkin-2017,Pagano-2018}.

\subsection{Initial condition} \label{sec:ic}

To combine the NLFFF with the MHD simulation the nonpotential, pre-eruptive magnetic field from the NLFFF simulation must be used as the initial configuration in the MHD simulation. To do this, we first import the components of the 3D magnetic field ($B_{r}, B_{\theta}, B_{\phi}$) from the NLFFF evolutionary model on the same day as the stealth event (day 123 of the simulation, 2015 January 3) into the MHD simulation and interpolate the components onto the numerical grid of the MHD simulation. The MHD equations are then solved numerically using PLUTO \citep{Mignone-2012}. However, the domain of the MHD simulation extends to $r = 4 R_{\odot}$ whereas, the domain of the NLFFF evolutionary model only extends to a source surface of $r=2.5 R_{\odot}$. Therefore, before performing the simulation we extended the magnetic field configuration taken from the NLFFF evolutionary model to the outer boundary of the MHD simulation. To extend the magnetic field configuration we assume a purely radial field beyond the original source surface at $r=2.5 R_{\odot}$
\begin{equation}
\label{brover25r}
B_r(r>2.5 R_{\odot},\theta,\phi)=B_r(2.5 R_{\odot},\theta,\phi)\frac{\left(2.5 R_{\odot}\right)^2}{r^2}.
\end{equation}

In order to produce a complete set of MHD variables for the initial condition we must also construct a distribution for the plasma density and temperature.

To begin with, when we imported the global magnetic field configuration, we adjusted our model to reproduce realistic values of plasma $\beta$, density, and temperature. Through doing this we aim to construct a plasma distribution that has the following key aspects of the density distribution of the solar corona: i) density and gravity stratification, ii) horizontal thermal pressure balance, and iii) an approximately uniform temperature distribution in quiet Sun regions. To achieve these key aspects in our simulation, we constructed the plasma distribution by starting from a simple static distribution of density $\rho$, and thermal pressure $p$. We then modified this static plasma distribution to differentiate regions of strong, highly concentrated magnetic field from quiet Sun magnetic field.
In the first step, the density of the background corona $\rho_\mathrm{b}$ follows a radial profile given by
\begin{equation}
\label{densitystratification}
    \rho_\mathrm{b} (r) = \rho_\mathrm{ph} \left(\frac{r}{R_\odot}\right)^{k},
\end{equation}
where $\rho_\mathrm{ph}$ is the photospheric density and $k$ is the power law index that describes the steepness of the density profile. The values of the photospheric density and power law index are given in Table~\ref{MHDparameters} in Sect.~\ref{initcond} along with the values of the remainder of the MHD parameters applied. The thermal pressure can then be defined by imposing that the atmosphere is in gravitational equilibrium as follows
\begin{equation}
\label{presstratification}
   \frac{\partial p}{\partial r} = - \frac{GM_{\odot}}{r^{2}}, 
\end{equation}
where $G$ is the gravitational constant and $M_{\odot}$ is the mass of the Sun. Equation~\ref{presstratification} can be solved analytically by using the boundary condition of the thermal pressure at the outer domain boundary
\begin{equation}
\label{pres4rsun}
    p_{4R_\odot} = \frac{\rho_\mathrm{b}(4R_{\odot})}{\mu m_\mathrm{p}} k_\mathrm{B}T_\mathrm{C},
\end{equation}
where $T_\mathrm{C}$ is the coronal temperature at $r=4 R_{\odot}$.

To account for regions of strong, highly concentrated magnetic field we use the proxy for magnetic flux rope formation $\omega(r,\theta,\phi)$, 
which has previously been defined in \citet{Rodkin-2017}, and \citet{Pagano-2018, Pagano-2019}, where
\begin{eqnarray}
\label{omegab}
\omega&=&\sqrt{\omega_r^2+\omega_{\theta}^2+\omega_{\phi}^2}, \\
\label{omegar}
\omega_r&=&\frac{\left|\vec{B}\times\vec{\nabla B_r}\right|}{\left|\vec{\nabla B_r}\right|}, \\
\label{omegatheta}
\omega_{\theta}&=&\frac{\left|\vec{B}\times\vec{\nabla B_{\theta}}\right|}{\left|\vec{\nabla B_{\theta}}\right|}, \\
\label{omegaphi}
\omega_{\phi}&=&\frac{\left|\vec{B}\times\vec{\nabla B_{\phi}}\right|}{\left|\vec{\nabla B_{\phi}}\right|}.
\end{eqnarray}

The function $\omega$ is positive definite and increases for complex and twisted magnetic field structures.
To visualise the physical meaning of the function $\omega$ consider a magnetic flux rope that lies above a PIL. The gradient of the vertical component of the magnetic field is horizontal and perpendicular to the PIL. In contrast, the magnetic field has a strong axial component that is horizontal and parallel to the PIL. As these two vectors are almost perpendicular this maximises the vector product in the numerator of Eq.\ref{omegar}-\ref{omegaphi}. Moreover, as $\omega$ is dependent on the magnetic field intensity its value is large at the solar surface and decreases with increasing heliocentric distance. We implement $\omega$ in the simulations through the following set of functions

\begin{eqnarray}
\Omega_B(\omega)&=&\frac{\arctan{\left(\frac{\omega-\omega^{\star}}{\Delta\omega}\right)}}{\pi}+0.5,\\
\Omega_\theta(\theta)&=&\frac{\arctan\left(\frac{\theta-(\pi-\theta^{\star})}{\Delta\theta}\right)}{\pi}-\frac{\arctan\left(\frac{\theta-\theta^{\star}}{\Delta\theta}\right)}{\pi}+1,\\
\Omega&=&\frac{\Omega_B+\Omega_{\theta}}{2}+\frac{\left|\Omega_B-\Omega_{\theta}\right|}{2},
\end{eqnarray}

where $\Omega_{\theta}$ and $\Omega_{B}$ are bound between 0 and 1, and $\Omega$ is defined to select the highest value between them. The transition between 0 and 1 depends upon the function parameters $\omega^{\star}$ and $\theta^{\star}$ (see Table~\ref{MHDparameters} in Sect.~\ref{initcond}). The purpose of $\Omega_{\theta}$ is to avoid boundary effects that occur near the polar regions. For more details see \citet{Pagano-2018}.

Using Eq.~\ref{densitystratification} we can now construct a temperature distribution $T$ from $\Omega$, which is defined as
\begin{equation}
\label{tempomega}
T = \Omega (T_\mathrm{FR} - T_\mathrm{C}) + T_\mathrm{C},
\end{equation}
where $T_\mathrm{FR}$ and $T_\mathrm{C}$ are the flux rope and coronal temperature, respectively. For highly concentrated magnetic fields $\Omega \sim$~1 whereas, $\Omega \sim$~0 for quiet Sun regions. 

To account for the newly defined temperature distribution, we modified the thermal pressure such that
\begin{equation}
\label{rhostep2}
\rho=\rho_\mathrm{b}\frac{T_\mathrm{C}}{T}.
\end{equation}
These plasma and temperature distributions result in a solar corona with cool, dense flux ropes in regions of strongly twisted magnetic field and hot, low density coronal arcades in the background corona. The final configuration is not in hydrostatic equilibrium as modifying the density distribution in Eq.~\ref{rhostep2} generates a force imbalance. However, the time scales over which plasma displacements that are induced by this imbalance occur are much longer than the time scales related to the Lorentz forces present in the domain.
In terms of the forces, we found that the unbalanced Lorentz force was about 10 times stronger than the pressure gradient and gravity at the locations where the eruptions are triggered, thus such departure from the hydrostatic equilibrium is not sufficient to prevent the eruption of unstable magnetic structures.

\subsection{MHD equations} 

As previously stated, once we have coupled the two techniques and defined the plasma and temperature distributions in the computational domain, we use PLUTO \citep{Mignone-2012} to solve the time-dependent MHD equations
\begin{align}
\label{mass}
&\frac{\partial\rho}{\partial t}+\vec{\nabla}\cdot(\rho\vec{v})=0,\\
\label{momentum}
&\frac{\partial\rho\vec{v}}{\partial t}+\vec{\nabla}\cdot(\rho\vec{v}\vec{v})
+\nabla p- \frac{1}{4\pi}(\vec{\nabla}\times\vec{B})\times\vec{B}=\rho\vec{g},\\
\label{induction}
&\frac{\partial\vec{B}}{\partial t}-\vec{\nabla}\times(\vec{v}\times\vec{B})=0,\\
\label{energy}
&\frac{\partial e}{\partial t}+\vec{\nabla}\cdot[(e+p)\vec{v}]=\rho\vec{g}\cdot\vec{v}.
\end{align}

The total energy density $\epsilon$ is given by
\begin{equation}
    \label{enercouple}
\epsilon = \frac{p}{\gamma-1}+\frac{1}{2}\rho\vec{v}^2+\frac{\vec{B}^2}{8\pi},
\end{equation}
where $\gamma=5/3$ is the ratio of specific heats. The expression for gravitational acceleration is given by
\begin{equation}
\label{solargravity}
\vec{g}=-\frac{G M_{\odot}}{r^2}\vec{\hat{r}},
\end{equation}
where $\vec{\hat{r}}$ is the unit vector.

The computational domain is composed of a 112 $\times$ 192 $\times$ 384 grid in the $r, \theta, \phi$ directions, with the cell size increasing in the radial direction from $\Delta = 0.012R_{\odot}$ at $r = 1R_{\odot}$, to $\Delta = 0.049R_{\odot}$ at $r = 4R_{\odot}$. The cell size in the $\theta$ and $\phi$ directions is uniform. The latitude $\theta$ and longitude $\phi$ span from $\theta = 0.75^{\circ}$ to $\theta= 179.25^{\circ}$, and 0 to 360$^{\circ}$, respectively. The boundary conditions were constructed using ghost cells in order to be consistent with the conditions from the NLFFF evolutionary model. In the simulation, the outer boundary is open, reflective boundary conditions are set for $\theta$, and the $\phi$ boundaries are periodic. 
At the lower photospheric boundary, we imposed fixed boundary conditions where the plasma and magnetic field cannot evolve below $r = 1.03 R_{\odot}$. As no driving is applied at the lower boundary in the MHD simulation, any dynamics produced were the result of the forces generated in the NLFFF model.



\subsection{MHD simulation initial condition} 
\label{initcond}
In order to model the stealth event that occurred in the observations on 2015 January 3, we imported the global magnetic field from the NLFFF evolutionary model on day 123 (see Figs.~\ref{fig:fig6} and \ref{fig:fig7} panel (a)) and followed the procedure outlined in Section~\ref{sec:ic}. At this time, a flux rope (FR3 in Fig.~\ref{fig:fig7}) has formed in the model at the location of the stealth event. Two other flux ropes F1 and F2 have also formed that are connected to the large AR complex on the farside.

\begin{table}
\caption{Parameters used in the MHD simulation.}      
\label{MHDparameters}    
\centering
\begin{tabular}{c c c}
\hline\hline
Parameter & Value & Units  \\
\hline 
   $\rho_{ph}$ & $2\times10^{-16}$ & g~cm$^{-3}$  \\      
   $k$ & -6 & \\
   $\theta^{\star}$ & $0.13$  & rads \\
   $\Delta\theta$ & $0.005$ & rads \\
   $\omega^{\star}$ & $3$  & G\\
   $\Delta\omega$ & $0.3$  & G \\
   $T_{FR}$ & $10^3$ & K \\
   $T_{C}$ & $10^6$ & K \\
\hline
\end{tabular}
\end{table}

\begin{figure*}
\centering
\includegraphics[width=1\textwidth]{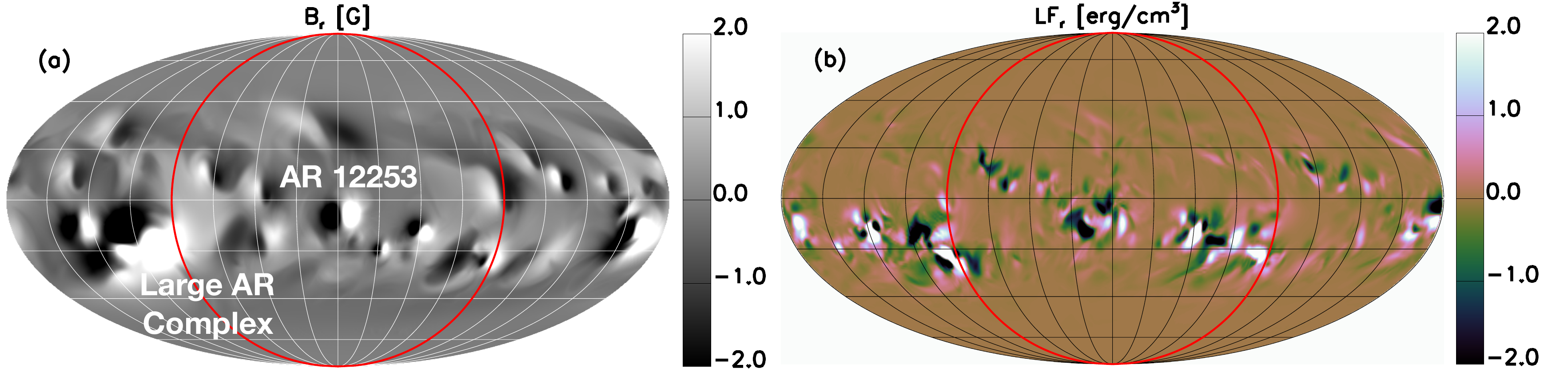}
\caption{Panel (a) shows the radial magnetic field component $B_{r}$ that is taken from the NLFFF evolutionary model and used as the lower boundary in the initial condition of the MHD simulation. The map of the surface magnetic field is shown using Mollweide projection where black (white) represents positive (negative) magnetic field regions. The region of the solar disk visible from SDO/Earth is represented by the red circles, AR 12253 and the large AR complex behind the east limb are labelled in white. Panel (b) is the radial component of the Lorentz force $LF_{r}$ at the solar surface.} 
\label{fig:fig8}
\end{figure*}

The parameter values used in the MHD simulation are given in Table~\ref{MHDparameters}. Figure~\ref{fig:fig8}(a) shows the radial magnetic field component, imported from the NLFFF evolutionary model, that is used as the lower boundary condition in the initial condition of the MHD simulation. The map shows that there are several active regions visible on disk during this time period that are also seen in the observations. In particular, we focus on the AR which is close to central meridian (AR 12253 in Fig.~\ref{fig:fig3}(a) and Fig.~\ref{fig:fig8}), the region directly below the AR, and the large complex of ARs (12235, 12237, and 12242) in Fig.~\ref{fig:fig3}(b)) located behind the east limb i.e. the region outside of the red circle in Fig.~\ref{fig:fig8}. The large AR complex is therefore not visible in the observations at the time of the eruption. Figure~\ref{fig:fig8}(b) shows the corresponding radial Lorentz force at the solar surface. The Lorentz force $LF_{r}$ is generally very small where the corona is close to equilibrium however, the values of $LF_{r}$ are roughly one order of magnitude larger at the locations that correspond to AR 12253 and the large AR complex.

\begin{figure*}
\centering
\includegraphics[width=1\textwidth]{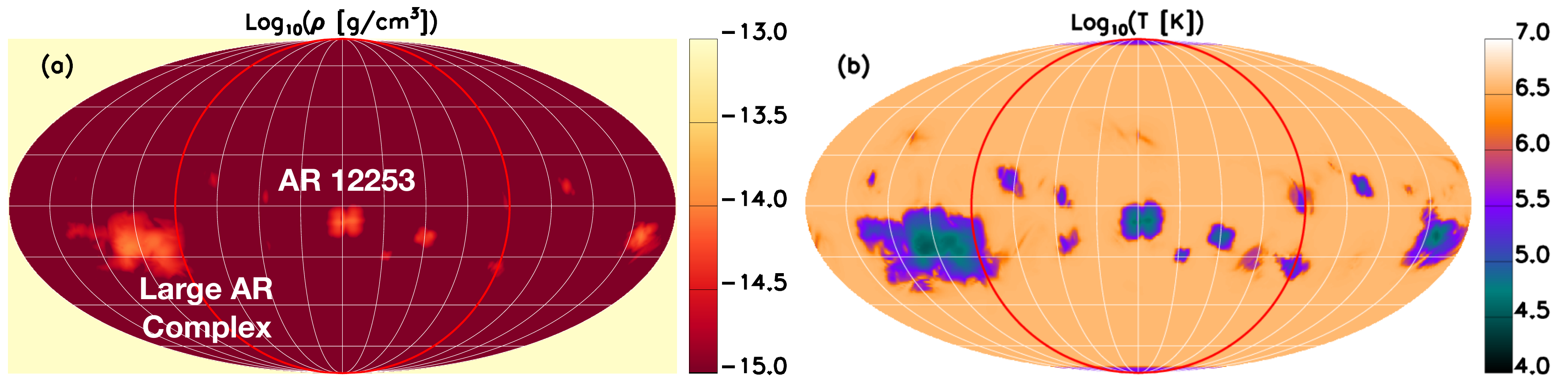}
\caption{Plasma (panel (a)) and temperature (panel (b)) distributions of the initial condition of the MHD simulation. The region of the solar disk visible from SDO/Earth is indicated by the red circles while AR 12253 located near central meridian and the large AR complex behind the east limb are labelled in white.}
\label{fig:fig9}
\end{figure*}

Panels (a) and (b) of Fig.~\ref{fig:fig9} show the distribution of density and temperature at the lower boundary of the initial condition of the MHD simulation. The active regions in the simulation are cold and dense compared to the surrounding hot, sparse solar corona. The initial condition of the simulation is in non-equilibrium as the magnetic field configuration imported from the NLFFF evolutionary model is not force-free as the build-up of stress and energy leads to the eruption of the flux ropes. As the solar atmosphere is constructed in such a way that the intensity of the downward-directed gravitational force is always equal to or larger than the outward-directed thermal pressure gradient, any upward plasma motion must be due to the presence of the Lorentz force in the initial condition.

\subsection{Evolution \& eruption dynamics}
\label{mhdevol}

\begin{figure*}
\centering
\includegraphics[width=1\textwidth]{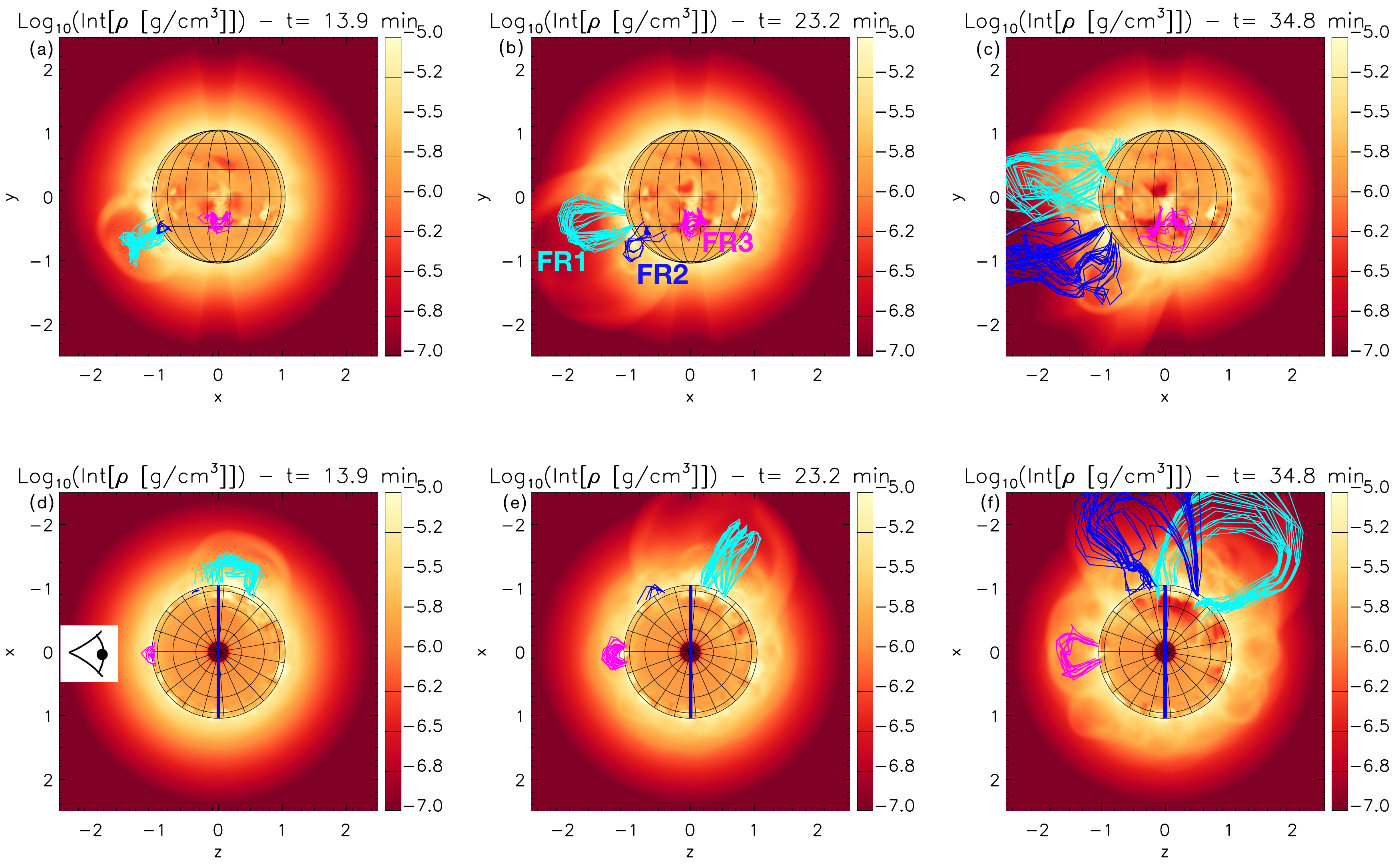}
\caption{Coronal density distribution integrated along the line-of-sight taken from the MHD simulation. The density is shown at two different vantage points (as viewed from Earth in panels (a)--(c) and above the solar north pole in panels (d)--(f)) at the times of $t = 13.9, 23.2,$ and $34.8~\mathrm{min}$. The blue line (bottom row) represents the plane-of-sky position. The eye symbol in the bottom row indicates the view as seen from SDO/at Earth. The panels also show the evolution of the magnetic field lines (light blue, dark blue, and magenta) of three flux ropes (FR1, FR2, and FR3) that erupt in the MHD simulation.}
\label{fig:fig10}
\end{figure*}

The coronal density distribution integrated along the line-of-sight is shown in Fig.~\ref{fig:fig10}, as viewed from the Earth (top row), and above the solar north pole (bottom row). The evolution of three flux rope structures (light blue, dark blue, and magenta field lines labelled FR1, FR2, and FR3, respectively) is also shown in Fig~\ref{fig:fig10}. These structures correspond to the three flux ropes identified in the NLFFF simulation and are associated with the most significant density perturbations in the MHD simulation. To give an estimation of the spatial size of the flux ropes the angular distance between the footpoints of the light blue, dark blue, and magenta flux ropes was calculated to be 11, 31, and 4$^{\circ}$, respectively.

At $t = 13.9~\mathrm{min}$ (Fig.~\ref{fig:fig10} panel (a)), there is clear evidence of an eruption taking place near the east limb. The eruption consists of a bright leading edge, dark cavity, and FR1. When viewed from above the Sun's north pole (panel (d)) the eruption originates from a region that is mostly behind the east limb. Taking into account the results from the observations, NLFFF evolutionary model, and the simulation, the eruption can be associated with the large AR complex (ARs 12235, 12237 and 12242). At this time, the coronal density elsewhere remains almost constant, except for the region associated with FR2, which is partially connected to the large AR complex. At $t = 13.9~\mathrm{min}$, FR2 has also just started to erupt. This second eruption of FR2 is more evident at subsequent timesteps (see Fig.~\ref{fig:fig10} panels (b) and (c)). 

In the final column of Fig.~\ref{fig:fig10}, at $t = 34.8~\mathrm{min}$, the density perturbation due to FR1 has reached the outer boundary of the simulation located at 4~R$_{\odot}$ while FR2 trails behind. At the same time, multiple eruptions occur on the solar disk however, only one has enough energy to leave the lower corona, and achieves a large enough radial distance to be distinguishable from the background corona. The structure responsible for the third eruption is FR3, the location of which is slightly offset from disk centre and to the south of AR 12253. The eruption of FR3 occurs at the same location as the coronal dimming in the observations and is therefore determined to be responsible for the stealth event.

\begin{figure*}
\centering
\includegraphics[width=1\textwidth]{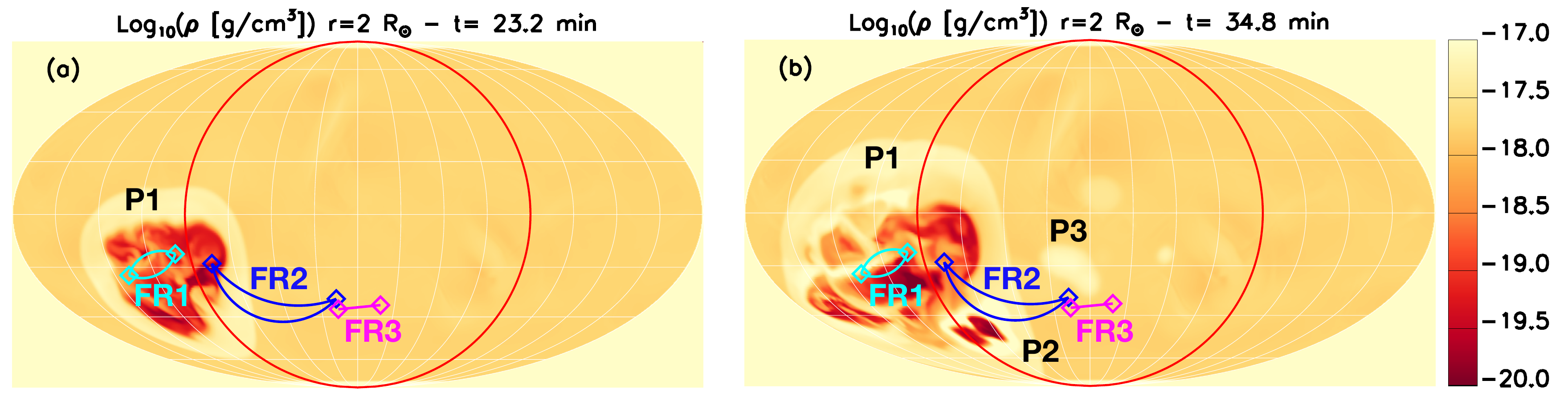}
\caption{Density distribution taken at $r = 2 R_{\odot}$ at $t = 23.2~\mathrm{min}$ and $t = 34.8~\mathrm{min}$ is shown in panels (a) and (b), respectively. Here, the red circles illustrate the portion of the Mollweide map that corresponds to the $\phi$ coordinates on the solar disk visible from the Earth vantage point. The density perturbations that are a result of three flux rope eruptions occurring in the MHD simulation are labelled P1, P2 and P3. The light blue, dark blue, and magenta structures represent the approximate locations of the three flux ropes (F1, F2, F3) and their footpoints taken from the NLFFF evolutionary model.}
\label{fig:fig11}
\end{figure*}

In order to investigate the propagation of the eruptions into interplanetary space we show the density distribution at $r = 2 R_{\odot}$ (Fig.~\ref{fig:fig11}) at $t = 23.2$ and $34.8$~min. The first density perturbation at $t = 23.2~\mathrm{min}$ (see P1 in Fig.~\ref{fig:fig11} panel (a)) covers a large circular area centred above the AR complex (ARs 12235, 12237 and 12242), which is located behind the east limb. The perturbation is a result of the eruption of the AR complex, which includes the ejection of FR1. The majority of the density enhancement due to this eruption originates from the farside, more than 90$^{\circ}$ away from the Sun-Earth line. This eruption would therefore not be Earth-directed or be connected to the stealth event. At a later time of $t = 34.8~\mathrm{min}$, there are two additional small perturbations in density (P2 and P3 in Fig.~\ref{fig:fig11} panel (b)) that reach the source surface, while the earlier density perturbation (P1) has expanded. The two perturbations located at the east limb and near disk centre (P2 and P3) are a result of the eruptions of FR2 and FR3. The first small density perturbation (P2), which is located at the east limb and within the original large density perturbation (P1), is due to the eruption of FR2 in the MHD simulation. The other additional small density perturbation (P3), located near disk centre, occurs later than the previous ones and is due to the eruption of FR3. As the eruption of FR3 is Earth-directed it has the potential to cause a geomagnetic storm at Earth. The sequence of events suggests that these eruptions could be sympathetic with the final density perturbation (P3), which is located close to disk centre, being a consequence of the coronal perturbations (P1 and P2) caused by the first and second flux rope (FR1 and FR2) eruptions.

\begin{figure*}
\centering
\includegraphics[width=1\textwidth]{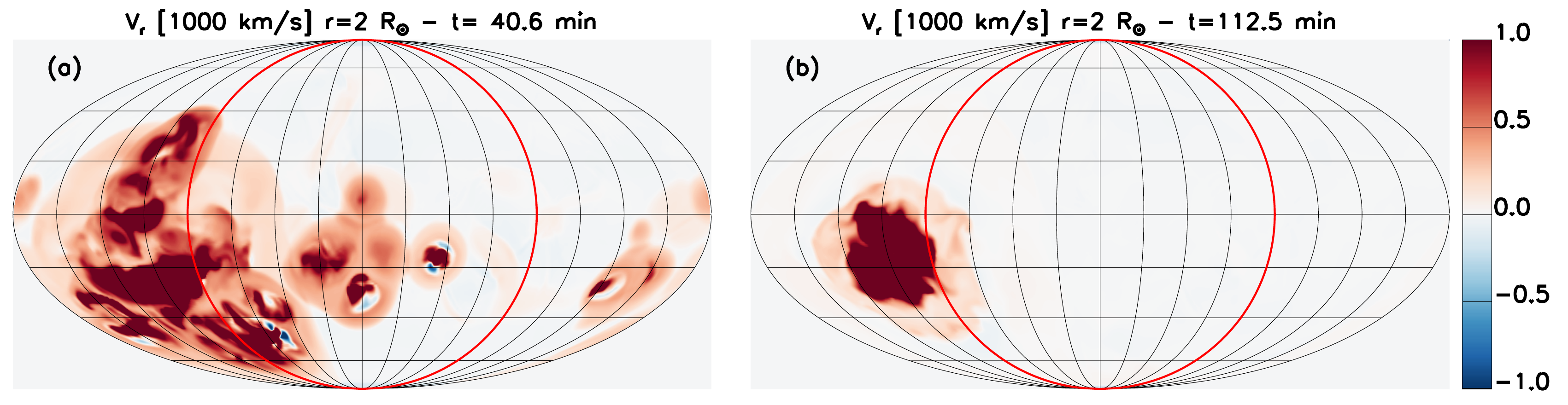}
\caption{Radial velocity at $r = 2 R_{\odot}$ for two different simulations where the photospheric density is varied. Panel (a) shows the radial velocity taken at $t = 40.6~\mathrm{min}$ for the simulation previously described where $\rho_\mathrm{ph} = 2\times10^{-16}~\mathrm{g~cm}^{3}$. Panel (b) shows the radial velocity from a second simulation at the later time of $t=112.5$~min, where the photospheric density has increased by an order magnitude $\rho_\mathrm{ph} = 2 \times 10^{-15}~\mathrm{g~cm}^{3}$. The red circle represents the solar disk which is visible at Earth. Red (blue) corresponds to positive (negative) velocities which are saturated at $\pm~1000~\mathrm{km~s}^{-1}$.} 
\label{fig:fig12}
\end{figure*}

In Fig.~\ref{fig:fig12} panel (a) we show the radial velocity $v_{r}$ at $r = 2 R_{\odot}$ at $t = 40.6~\mathrm{min}$ in the MHD simulation. It is evident that the three flux rope eruptions that occur on the farside (FR1), at the east limb (FR2), and south of disk centre (FR3) generate fast plasma outflows. The simulation also shows some small localised downflows at the periphery of the eruptions of FR2, and FR3. These downflows are likely due to small amounts of plasma that fall back to the surface during the eruption process.

We note that the photospheric density and subsequently the derived coronal values can significantly affect the MHD evolution in the model and that the outflows encountered in our model are significantly higher than the ones measured by LASCO. Therefore, to understand the full implications of the atmospheric reconstruction on the eruption dynamics, and in order to reduce the discrepancies in the observed speeds, we perform a second simulation where we change the value of photospheric density in the model. In the second simulation we set $\rho_{ph} = 2 \times 10^{-15}~\mathrm{g~cm}^{3}$, compared to a density of $2 \times 10^{-16}~\mathrm{g~cm}^{3}$ used in the first model. The radial velocity distribution produced by the second simulation is shown in panel (b) of Fig.~\ref{fig:fig12}. In this simulation, we find that only the largest eruption from the AR complex of FR1 has produced significant plasma outflows in the solar corona whereas, the rest of the corona remains in equilibrium. A variation in photospheric density of a factor 10 is not unusual and similar results were found in \citet{Pagano-2013b}. In this current study it is shown that a successful eruption is dependent on both the density distribution of the atmosphere and the magnetic field configuration. Moreover, these results show that many eruption scenarios are possible and that observational constraints are crucial in determining the correct scenario.
At the same time, in this model the plasma $\beta$ is proportional to the density, which leads to a less accurate description of the coronal physics when $\beta$ grows above $\sim1$. Values of $\beta$ slightly higher than coronal ones, but still lower than 1, are necessary to allow the plasma to re-adjust when the NLFFF configuration is imported. We identify this as an important area for improvement, where in future studies, we aim to import the NLFFF in two steps to allow for a more accurate description of the solar atmosphere in terms of plasma density, plasma $\beta$, and eruption outflows.


\section{Discussion \& conclusions}
\label{sec:sum}

In this study, we have coupled the nonlinear force-free field (NLFFF) evolutionary model with a global MHD simulation in order to investigate the origin and dynamics of the stealth event that occurred on 2015 January 3. The NLFFF evolutionary magnetic field model provides an accurate description of the global magnetic field of the solar corona in the months prior to eruption onset. We used the global magnetic field from the NLFFF evolutionary model taken on the same day as the occurrence of the stealth event in the observations as the initial condition of the global MHD simulation. To obtain a full set of variables for the MHD simulation, we constructed distributions for the plasma density and temperature. We then use PLUTO \citep{Mignone-2012} to solve the time-dependent MHD equations in the computational domain. 
By coupling the two techniques we propose the following scenario that successfully describes the occurrence and signatures of the stealth event, the CME observed in LASCO/C2, and the geomagnetic disturbance present in the remote and in-situ observations. Two magnetic flux ropes (FR1 and FR2 in Fig.~\ref{fig:fig10} panel (b)) erupt first producing two near-simultaneous CMEs both directed away from Earth. A third, weaker eruption follows that corresponds to the eruption of a flux rope structure (FR3 in Fig.~\ref{fig:fig10} panel (b)) that is south of AR 12253 and north of the southern polar coronal hole. The weak eruption due to FR3 is responsible for the geomagnetic storm at Earth.

In the MHD simulation, the third flux rope eruption originates from a structure which is less unstable than the first two. This scenario agrees with previous studies that suggest that stealth CMEs are the result of sympathetic eruptions and the reconfiguration and removal of the overlying, stabilizing magnetic field can trigger a neighbouring CME (e.g. \citealt{OKane-2019}).

The scenario put forward by the results from the coupled simulations is supported by the observations as follows. In the lead up to the eruption of the stealth event, a large AR complex, composed of ARs 12235, 12237, and 12242, is visible on the solar disk and rotates around to the farside of the Sun. This large AR complex is located behind the east limb at the time of the stealth event and is therefore not visible in the coronal observations at the time of the stealth eruption.

If we revisit the LASCO observations a single partial halo CME is recorded at around 03:12~UT on 2015 January 3. When examining the LASCO/C2 observations in more detail the leading edge of the CME appears to consist of the combination of two CME fronts where one is roughly $1~R_{\odot}$ ahead of the other. Taking the velocity of the first CME to be around $500~\mathrm{km~s}^{-1}$, this corresponds to a time difference of $\sim$30~mins between the two eruptions, which is consistent with the first triggering the second eruption given an Alfv\'en speed of $1000~\mathrm{km~s}^{-1}$ and that they are roughly half a radian apart on the solar disk. A possible explanation for this, considering the results from the MHD simulation (see Fig.~\ref{fig:fig10}), is that the light blue flux rope (FR1 in Fig.~\ref{fig:fig10}) associated with the large AR complex erupts first, immediately followed by the eruption of the dark blue flux rope (FR2 in Fig.~\ref{fig:fig10}), that is partially connected to the large AR complex. In the MHD simulation, the density enhancements (P1 and P2 in Fig.~\ref{fig:fig11}) due to the propagation of these eruptions (FR1 and FR2) into space (Fig.~\ref{fig:fig11}) are mainly visible behind or at the east limb. Therefore, the eruptions of FR1 and FR2 would not be Earth-directed. Due to the optically thin nature of the observations and the two eruptions occurring almost simultaneously, leads to the interpretation that a single partial halo CME occurred in the coronagraph observations. Unfortunately, there are no STEREO observations available at this time to observe the two CMEs from a different perspective. 
 
A few days later on 2015 January 7, the signatures of a magnetic cloud are observed in the in-situ data, which causes a strong geomagnetic storm at Earth. While it is difficult to determine the source region, a faint coronal dimming visible in the SDO/AIA 211~\AA\ observations is thought to be the source of the magnetic cloud. This is corroborated by the presence of a flux rope structure (FR3 in Fig.~\ref{fig:fig7}) in the NLFFF model at the same location of the coronal dimming in the observations. This flux rope located to the south of AR 12253 erupts in the MHD simulation after FR1 and FR2 (see Fig.~\ref{fig:fig10}). We calculated the angular separation of the footpoints of the magenta flux rope (FR3) in the simulation to be around 4$^{\circ}$, which is much smaller than the other two erupting flux ropes (11 and 31$^{\circ}$). Given the size, location, and the very weak eruptive signatures observed in the low corona associated with the eruption meant that this CME wasn't visible in LASCO and the resulting geomagnetic disturbance was attributed incorrectly to the CME observed in LASCO/C2 by space weather forecasters. These observations and simulation results highlight the importance of an operational L5 mission for space weather forecasting.

Although the faint coronal dimming visible on the disk in the 211\,\AA\ observations is the most probable source of the stealth event other sources cannot be ruled out. \citet{Cid-2016} discusses several possible sources including several filament candidates that could be responsible for the stealth event (see Sect.~\ref{sec:obs}). \citet{Nitta-2017} present the possibility that the coronal dimming is the cause however, they also discuss the possibility that an area beyond AR 12253 could be involved in the eruption. This is possible and confirmed by the density distribution and radial velocity taken from the MHD simulation at $r = 2 R_{\odot}$. The density structure (Fig.~\ref{fig:fig10} (b)) is slightly offset from the location of the flux rope structure in the NLFFF evolutionary model. This offset could be due to deflection of the flux rope as it propagates through the corona caused by the open magnetic field of the coronal hole. This made the determination of the region primarily responsible for the stealth event difficult.

Furthermore, the coupling between the NLFFF model and MHD simulation involves an atmospheric reconstruction that could be better constrained by the observations. In future studies we plan to use advanced diagnostic techniques that can constrain our models using observations. In particular, this will be possible with the new generation of coronagraphs, such as METIS \citep{Antonucci-2017} on board {\it Solar Orbiter} \citep{Muller-2020}. For example, white-light observations can be used to constrain the column density variations during a CME. These constraints are more accurate when polarised light observations are also available \citep{Bemporad-2015,Pagano-2015}. In addition, it would be possible to constrain the temperature distribution using Lyman-$\alpha$ observations as outlined in \citet{Bemporad-2018} and the plane-of-the-sky velocity field \citet{Ying-2019}. The present study does however, illustrate the power in using a coupled data-driven NLFFF and MHD approach to model eruptions on the Sun. Through this we have been able to determine the origin of both the LASCO CME and a stealth CME, and shown that while the two are related, they have different origin locations on the Sun.


\begin{acknowledgements}
      S.L.Y. would like to acknowledge STFC for support via the consolidated grant SMC1/YST037 and also NERC for funding via the SWIMMR Aviation Risk Modelling (SWARM) Project, grant number NE/V002899/1. P.P. would like to thank the ERC for support via grant No. 647214. D.H.M. would like to thank the STFC for support via consolidated grant ST/N000609/1 and, the Leverhulme trust, and the ERC under the Synergy Grant: The Whole Sun (grant agreement no. 810218) for financial support. P.P. and D.H.M. would like to thank STFC for IAA funding under grant number SMC1-XAS012. L.A.U. was supported by the NSF Atmospheric and Geospace Sciences Postdoctoral Research Fellowship Program (Award AGS-1624438). We would also like to thank the ISSI Team led by Nariaki Nitta and Tamitha Mulligan for the useful discussions on Problem Geomagnetic Storms. 
      This research used version 1.0.2 (\url{https://doi.org/10.5281/zenodo.3779284}) of the SunPy open source software package \citep{sunpycommunity2020} and also JHelioviewer \citep{Muller-2017}. The OMNI data were obtained from the GSFC/SPDF OMNIWeb interface at omniweb.gsfc.nasa.gov. We also used data taken from the LASCO CATALOG (cdaw.gsfc.nasa.gov/CME\_list/). This work used the DiRAC@Durham facility managed by the Institute for Computational Cosmology on behalf of the STFC DiRAC HPC Facility (www.dirac.ac.uk). The equipment was funded by BEIS capital funding via STFC capital grants ST/P002293/1, ST/R002371/1 and ST/S002502/1, Durham University and STFC operations grant ST/R000832/1. PLUTO was developed at the Turin Astronomical Observatory in collaboration with the Department of Physics of the Turin University.
      
\end{acknowledgements}

\bibliographystyle{aa} 
\bibliography{ref.bib} 

\end{document}